\documentclass[sigconf, natbib, screen]{acmart}

\usepackage{multirow}
\usepackage{mathtools}
\usepackage{svg}
\usepackage[font=small,skip=0pt]{caption}
\usepackage{graphicx}
\usepackage{subcaption}
\usepackage[inline]{enumitem}
\usepackage{booktabs}
\usepackage{adjustbox}
\usepackage{hyphenat}
\usepackage{acronym}
\usepackage{cleveref}[2012/02/15]
\crefformat{footnote}{#2\footnotemark[#1]#3}
\usepackage{pifont}
\newcommand{\cmark}{\ding{51}}%
\newcommand{\xmark}{\ding{55}}
\usepackage{bigdelim}

\makeatletter
\renewcommand\paragraph{\@startsection{paragraph}{4}{\z@}%
                       {-2\p@ \@plus -1\p@ \@minus -1\p@}%
                       {-0.5em \@plus -0.22em \@minus -0.1em}%
                       {\normalfont\normalsize\itshape}}
\makeatother

\newcommand{\header}[1]{\vspace*{0.2mm}\paragraph{\bf #1.}}

\acrodef{L2R}{learning-to-rank}
\acrodef{SERP}{search result page}
\acrodef{nDCG}{normalized discounted cumulative gain}
\acrodef{PDS}{policy distributional shift}
\acrodef{OPE}{off-policy evaluation}

\copyrightyear{2023}
\acmYear{2023}
\setcopyright{acmlicensed}
\acmConference[WSDM '23]{Proceedings of the Sixteenth ACM International Conference on Web Search and Data Mining}{February 27-March 3, 2023}{Singapore, Singapore}
\acmBooktitle{Proceedings of the Sixteenth ACM International Conference on Web Search and Data Mining (WSDM '23), February 27-March 3, 2023, Singapore, Singapore}
\acmPrice{15.00}
\acmDOI{10.1145/3539597.3570412}
\acmISBN{978-1-4503-9407-9/23/02}



\author{Romain Deffayet}
\orcid{0000-0001-8265-9092}
\affiliation{%
\institution{Naver Labs Europe}
\city{Meylan}
\country{France}}
\affiliation{%
\institution{University of Amsterdam}
\city{Amsterdam}
\country{The Netherlands}}
\email{romain.deffayet@naverlabs.com}

\author{Thibaut Thonet}
\orcid{0000-0003-0302-0376}
\affiliation{%
\institution{Naver Labs Europe}
\city{Meylan}
\country{France}}
\email{thibaut.thonet@naverlabs.com}

\author{Jean-Michel Renders}
\orcid{0000-0002-7516-3707}
\affiliation{%
\institution{Naver Labs Europe}
\city{Meylan}
\country{France}}
\email{jean-michel.renders@naverlabs.com}

\author{Maarten de Rijke}
\orcid{0000-0002-1086-0202}
\affiliation{%
\institution{University of Amsterdam}
\city{Amsterdam}
\country{The Netherlands}}
\email{m.derijke@uva.nl}

\title{Generative Slate Recommendation with Reinforcement Learning}

\begin{abstract}
Recent research has employed reinforcement learning (RL) algorithms to optimize long-term user engagement in recommender systems, thereby avoiding common pitfalls such as user boredom and filter bubbles. They capture the sequential and interactive nature of recommendations, and thus offer a principled way to deal with long-term rewards and avoid myopic behaviors. However, RL approaches are intractable in the slate recommendation scenario -- where a list of items is recommended at each interaction turn~-- due to the combinatorial action space. In that setting, an action corresponds to a slate that may contain any combination of items. 

While previous work has proposed well-chosen decompositions of actions so as to ensure tractability, these rely on restrictive and sometimes unrealistic assumptions. Instead, in this work we propose to encode slates in a continuous, low-dimensional latent space learned by a variational auto-encoder. Then, the RL agent selects continuous actions in this latent space, which are ultimately decoded into the corresponding slates. By doing so, we are able to \begin{enumerate*}[label=(\roman*)] \item relax  assumptions required by previous work, and \item improve the quality of the action selection by modeling full slates instead of independent items, in particular by enabling diversity. \end{enumerate*} Our experiments performed on a wide array of simulated environments confirm the effectiveness of our generative modeling of slates over baselines in practical scenarios where the restrictive assumptions underlying the baselines are lifted.
Our findings suggest that representation learning using generative models is a promising direction towards generalizable RL-based slate recommendation.
\end{abstract}

\begin{document}		

\maketitle

\acresetall

\vspace{-3mm}

\section{Introduction}
\label{intro}

Ubiquitous in online services, recommender systems (RSs) play a key role personalization by catering to users' identified tastes. Ideally, they also diversify their offerings and help users discover new interests \cite{Jannach2021}. 
In the latter case, RSs take on an active role, which means that recommendations influence future user behavior, and therefore their effects on users must be explicitly controlled.
Such effects can be detrimental: users may get bored if too many similar recommendations are made, and it has been well-documented that users can end up in so-called filter bubbles or echo chambers \citep{Pariser2011,Bakshy2015,Flaxman2016}.
From the perspective of the online platform or the content provider, user boredom leads to poor retention and conversion rates \citep{long-term-better}, while filter bubbles raise fairness and ethical issues for which providers can be held accountable \citep{Masrour2020}. Conversely, RSs can also positively impact users, for example, when users get interested in new, unexpected topics or when the RS offers a fair representation of available options \citep{spotify-div}. 
%
It is natural, therefore, to balance exploitation (i.e., sticking to the known interests of the user) and exploration (i.e., further probing the user's interests) so as to avoid always recommending similar items, and encourage recommendations that boost future engagement. The reinforcement learning (RL) literature has proposed models and algorithms that aim to optimize long-term metrics by acknowledging the causal effect of recommendations on users \citep{youtube-topk, rl-longterm}.

In this work we consider the common scenario of slate recommendation \citep{wolpertinger,slateQ,youtube-topk}, which comes with specific challenges. At each interaction turn, a slate recommender system recommends a list of items from the collection, and the user interacts with zero, one or several of those items. As a consequence, users may not examine all the recommended items, which leads to biases in the observed interactions along with a complex interplay between items in the same slate~\citep{RIPS}. More importantly, the size of the action space, i.e., the number of possible slates, prohibits the use of off-the-shelf RL approaches~\citep{DulacArnold2015}. Indeed, as slate recommendation is a combinatorial problem, the evaluation of all actions by the RL agent through trial and error is simply intractable: even with as few as $1,000$ items in the collection, the number of possible slates of size $10$ is approximately $9.6 \times 10^{29}$. We propose to tackle this problem in the context of a practical scenario, \textbf{(S)}, which fits the second-stage ranking phase \citep{VanDang13} of many content recommendation platforms:
\begin{itemize}[label=(\textbf{S})]
\item \emph{The collection contains around a thousand items, and at each turn of interaction the proposed model must select and rank 10 items to be presented to the user.} 
\end{itemize}
All our tractability and feasibility statements in this paper must therefore be understood through the lens of this scenario \textbf{(S)}.

To reduce the prohibitively large size of the combinatorial action space, previous studies have proposed to decompose slates in a tractable manner \cite{wolpertinger,slateQ,youtube-topk}~-- but at the cost of restrictive assumptions, e.g., concerning mutual independence of items in the slate, knowledge of the user click model, availability of high-quality item embeddings, or that at most one item per slate is clicked. 

In contrast, in this work we propose to first learn a continuous, low-dimensional latent representation of actions (i.e., slates), and then let the agent take actions within this latent space during its training phase. In practice, we obtain the latent representations by introducing a \emph{generative modeling of slates} (GeMS) based on a variational auto-encoder (VAE) pre-trained on a dataset of observed slates and clicks, collected from a previous version of the recommender system. Such a dataset is usually available in industrial recommendation settings. Therefore, we do not rely on restrictive assumptions, and the fact that we represent full slates enables the agent to improve the quality of its recommendations, instead of using individual item representations. 

Our contributions can be summarized as follows:
\begin{itemize}[leftmargin=*,nosep]
    \item We propose GeMS, a novel way to represent actions in RL for slate recommendation, by pre-training a VAE on slates and associated clicks. Unlike previous methods, GeMS is free of overly restrictive assumptions and only requires logged interaction data.
    \item We provide a unified terminology to classify existing slate recommendation approaches based on their underlying assumptions.
    \item We show on a wide array of simulated environments that previous methods underperform when their underlying assumptions are lifted (i.e., in practical settings), while GeMS allows us to recover highly rewarding policies without restrictive assumptions.
    \item To support the reproducibility of this work, we publicly release the code for our approach, baselines and simulator.\footnote{\href{https://github.com/naver/gems}{https://github.com/naver/gems}.}
\end{itemize}


\section{Related Work}
\label{related-work}

\header{Long-term user engagement} Several studies have documented the misalignment between short-term benefits and long-term user engagement \citep{spotify-div, long-term-better}, as well as the tendency of traditional recommender systems to be detrimental to long-term outcomes \citep{frasca-closedloop}. Such myopic behavior is known to cause boredom and decrease user retention \citep{spotify-div}, which is prejudicial for both users and content providers. This behavior also raises concerns such as the rich-get-richer issue \citep{youtube-topk} and feeding close-mindedness \citep{frasca-closedloop}. Some previous studies tried to counter this effect by explicitly maximizing diversity \citep{spotify-GS} or by finding metrics correlated with long-term outcomes \citep{survival-models, surrogate-index}. 
In contrast, in our work we directly optimize long-term metrics by using reinforcement learning algorithms \citep{youtube-topk, rl-longterm, RL-diversity}. 

\header{Reinforcement learning for slate recommendation} The problem of slate recommendation with reinforcement learning (RL) has been tackled in several previous studies, although the settings in which solutions were tested vary and are sometimes not applicable to our scenario \textbf{(S)}. \citet{youtube-topk} and \citet{IRecGAN} assume a simple user click model and independence of items within a slate in order to reduce the problem to choosing individual items, which they solve with the REINFORCE algorithm on a SoftMax policy. \citet{slateQ} assume knowledge of the user's click model and item relevance, which allows them to perform combinatorial optimization for the computation of Q-values. \citet{wolpertinger} take a continuous action in the product space of item embeddings, i.e., one embedding per slot in the slate, and pre-select nearest-neighbor items for full-slate Q-function evaluation. \citet{cascading-Q} use properties of the optimal Q-function to propose an elegant decomposition of it and generate optimal slates autoregressively. We detail the assumptions made by each of these approaches in Section \ref{baselines}, but we had to discard \citep{cascading-Q} due to its prohibitively heavy computation: it requires a number of neural network forward passes proportional to the slate size times the number of items in the collection (i.e., 10,000 passes in scenario \textbf{(S)}), for each training or inference step. 

Our proposed approach differs from previous work because we do not manually decompose the slates using tractable heuristics based on restrictive assumptions, but instead approximate the slate generation process with a deep generative model.
Our proposed framework only has a single requirement, viz.\ the availability of logged data with slates and associated clicks, as we will detail in Section~\ref{baselines}. The latter assumption is by no means restrictive as such logged data is readily available in common industrial recommendation settings.

\header{Latent action representations} While learning a latent representation of states is very common in the RL literature \citep{representation-RL, world_models}, few studies have tackled the problem of latent action representation. \citet{AC-RA} train an action generation function in a supervised manner, by learning to predict the action taken from a pair of successive states. This is not directly applicable in our case, because the true user state is not observable and successive observations are simply clicks that appear to be too weak of a signal to infer the slates leading to these clicks. \citet{SA-WM} learn a state-action world model and jointly train latent state and action representations in a model-based fashion. 

\begin{figure*}[!ht]
\centering
\includegraphics[width=0.98\textwidth]{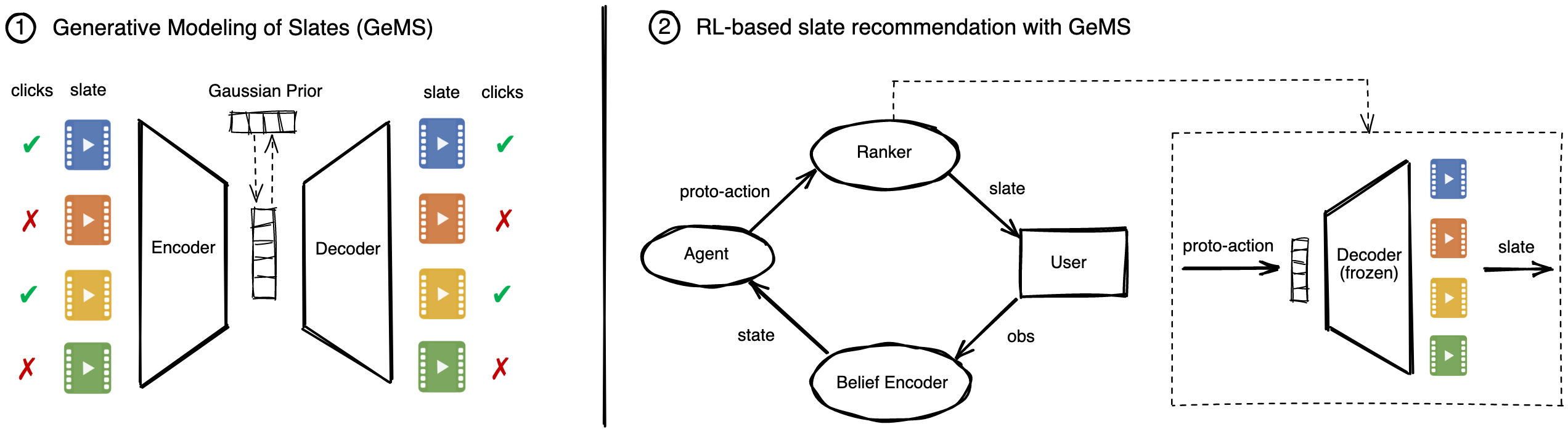}
\vspace{0.2cm}
\caption{Our proposed framework for slate recommendation with reinforcement learning. We first pretrain our GeMS model on previously collected logged data composed of slates and associated clicks (left), then we use the frozen decoder of GeMS to decode the RL agent's low-dimensional proto-action vector into a slate (right).}
\label{fig:SVAE_framework}       
\end{figure*}

Learning a world model in our setting essentially amounts to the latent modeling of slates and clicks (similar to our approach), while also conditioning on an internal hidden state.\footnote{We tried a similar method in pilot experiments, but the additional conditioning only deteriorated the results, so we only present the condition-free method in this paper.} The work by \citet{PLAS} is perhaps the closest work to ours, as it uses a variational auto-encoder (VAE) to embed actions into a controllable latent space before training an RL agent. However, it does not consider slates but only simple, atomic actions. In contrast, \citet{List-CVAE, Pivot-CVAE} train VAEs to represent slates and their associated clicks, but they do not investigate training an RL agent from the learned latent representation. 

To the best of our knowledge, we are the first to learn a latent representation of slates for RL-based recommendation.

\vspace{-3mm}

\section{Method}
\label{method}

\subsection{Notations and problem definition}
\label{problem-definition}

We consider a slate recommendation scenario in which a user interacts with a recommender system (RS) throughout an episode of $T$ turns. At every turn $t \in \{1, \dots, T\}$, the system recommends a slate $a_t = (i_t^1, \dots, i_t^k)$ where $(i_t^j)_{1 \leqslant j \leqslant k}$ are items from the collection $\mathcal{I}$ and $k$ is the size of the slate set by the RS designer. The user can click on zero, one or several items in the slate and the resulting click vector $c_t = (c_t^1, \dots, c_t^k), c_t^j \in \{0, 1\}$ is returned to the RS. 

The problem of maximizing the cumulative number of clicks over an episode can be modeled as a partially observable Markov decision process (POMDP) $\mathcal{M}^P = (\mathcal{S}, \mathcal{O}, \mathcal{A}, R, T, \Omega)$ defined by:
\begin{itemize}[leftmargin=*, nosep]
    \item A set of states $\mathcal{S}$, which represent the unobservable state of the user's mind;
    \item A set of observations $\mathcal{O}$ accessible to the system. Here, observations are clicks from the previous interaction ($o_t = c_{t-1}$) and therefore lie in the space of binary vectors of size $k$: $\mathcal{O} = \{0, 1\}^k$;
    \item A set of actions $\mathcal{A}$, which is the set of all possible slates composed of items from the collection, i.e., $|\mathcal{A}| = \frac{|\mathcal{I}|!}{(|\mathcal{I}| - k)!}$;
    \item A reward function $R : \mathcal{S} \times \mathcal{A} \rightarrow \mathbb{R}$, which we set to $R(s_t,a_t) = r_t = \sum_{j=1}^k c_t^j$ in order to reflect our long-term objective of maximizing the cumulative number of clicks; and
    \item A set of unknown transition and observation probabilities, respectively $T : \mathcal{S} \times \mathcal{A} \times \mathcal{S} \rightarrow [0,1]$ and $\Omega : \mathcal{S} \times \mathcal{A} \times \mathcal{O} \rightarrow [0,1]$, as well as a distribution over initial states $S^1 : \mathcal{S} \rightarrow [0,1]$.
\end{itemize}
\noindent%
Due to the unobserved nature of the true user state in the POMDP, it is common to train agents by relying on a proxy of the state inferred from available observations. The function that provides such proxy is traditionally referred to as the \textit{belief encoder} \citep{belief}. We also define the concepts of a policy $\pi : \mathcal{S \times A} \rightarrow [0,1]$ and trajectory $\tau = (o_t, a_t, r_t)_{1 \leqslant t \leqslant T}$. In the remainder, we write $\tau \sim \pi$ to signify that we obtain a trajectory by first sampling an initial state $s_1$ from $S^1$ and then recursively sampling actions $T -1$ times from the policy $\pi$. The goal can now be formulated as finding an optimal policy, i.e., a policy maximizing the \emph{expected return} $\pi^* \in \mathrm{arg}\max_\pi \mathbb{E}_{\tau \sim \pi} \left[ \mathcal{R}(\tau) \right]$ with $\mathcal{R}(\tau) = \sum_{t=1}^T r_t$. Finally, given a state $s$ and action $a$, we define the Q-function $Q^\pi(s,a) = \mathbb{E}_{\tau \sim \pi, s_1 = s, a_1 = a} \left[ \mathcal{R}(\tau) \right]$ and V-function $V^\pi(s) = \mathbb{E}_{a \sim \pi(s)} \left[ Q^\pi(s,a) \right]$.

\subsection{Overview of the framework}
\label{framework}

In our proposed framework, the interactions with the environment, i.e., the user, can be described by the following repeated steps:
\begin{enumerate}[leftmargin=*]
    \item The \emph{belief encoder} summarizes the history of interactions with the user into a state vector;
    \item The \emph{agent} selects a proto-action based on this state; and
    \item The \emph{ranker} (here resulting from a VAE model) decodes this proto-action into a slate that is served to the user.
\end{enumerate}

\noindent%
In the remainder of this section, we first detail our proposed \emph{generative modeling of slates} (GeMS). GeMS is a deep generative model that learns a low-dimensional latent space for slates and associated clicks~-- thus constituting a convenient proto-action space for the RL agent and allowing for tractable RL without resorting to restrictive assumptions as in prior work~\citep{youtube-topk, IRecGAN, slateQ, wolpertinger}. Then we describe how GeMS is integrated as a ranker in our RL framework and we briefly discuss the remaining RL components. This two-step process is depicted in Figure \ref{fig:SVAE_framework}.


\subsection{Generative Modeling of Slates (GeMS)}
\label{gems}

In order to instantiate our GeMS model, we propose to train a variational auto-encoder (VAE) on a precollected dataset $\mathcal{D}$ of logged interactions, as illustrated in Figure~\ref{fig:SVAE_framework} (left). 
A VAE aims to learn a joint distribution over data samples (i.e., slates and clicks denoted as $a$ and $c$, respectively) and latent encodings (i.e., proto-actions denoted as $z$) \citep{VAE}. To do so, a parameterized distribution $p_\theta(a,c,z)$ is trained to maximize the marginal likelihood of the data $p_\theta(a,c) = \int_z p_\theta(a,c,z) dz$.
In practice, due to the intractability of this integral, a parameterized distribution $q_\phi(z|a,c)$ is introduced as a variational approximation of the true posterior $p_\theta(z|a,c)$ and the VAE is trained by maximizing the evidence lower bound (ELBO):
$$\mathcal{L}_{\theta, \phi}^{\mathrm{ELBO}} \!=\! \mathbb{E}_{a,c \sim \mathcal{D}} \left[  \mathbb{E}_{z \sim q_\phi(\cdot | a,c)} \!\left[ \log p_\theta(a,c|z) \right] \!-\! \mathrm{KL} \left[ q_\phi(z|a,c) \| p(z) \right]  \right]\!,$$ where $p(z)$ is the prior distribution over the latent space, $\mathrm{KL}$ is the Kullback-Leibler divergence \citep{KL}, and $z$ is a sample from a Gaussian distribution obtained using the reparameterization trick \citep{VAE}. The distributions $q_\phi(z|a,c)$ and $p_\theta(a,c|z)$ are usually referred to as the encoder and the decoder, respectively.

The downstream performance of the RL agent we wish to ultimately learn clearly depends on the upstream ability of the VAE to properly reconstruct slates. However, as \citet{Pivot-CVAE} observe, an accurate reconstruction of slates may limit the agent's capacity to satisfy the user's interests. Indeed, finding high-performance continuous control policies requires smoothness and structure in the latent space, which may be lacking if too much emphasis is given to the reconstruction objective in comparison to the prior matching objective enforced by the KL-divergence. Therefore, it is necessary to balance reconstruction and controllability, which is done by introducing an hyperparameter $\beta$ as weight for the KL term in Eq.~\ref{eq:vae}. Moreover, in order to promote additional structure in the latent space, we add a click reconstruction term in the loss: slates with similar short-term outcomes (i.e., clicks) are grouped together during pre-training. Yet, we may want to avoid biasing the learned representations towards click reconstruction too much, as it may come at the cost of quality of the slate reconstruction. Therefore, we introduce a hyperparameter $\lambda$ to adjust this second trade-off. We show the empirical impact of $\beta$ and $\lambda$ in Section \ref{exp:tradeoffs}.
 
In our implementation, the prior $p(z)$ is set as a standard Gaussian distribution $\mathcal{N}(\mathbf{0}, \mathbf{I})$. The encoder $q_\phi(z |a, c)$ is a Gaussian distribution with diagonal covariance \smash{$\mathcal{N}(\mu_\phi(a,c), \text{diag}(\sigma^2_\phi(a,c)))$}, parameterized by a multi-layer perceptron (MLP). This MLP inputs the concatenation of learnable item embeddings and associated clicks over the whole slate, and outputs \smash{$(\mu_\phi(a,c), \log \sigma_\phi(a,c))$}. For the decoder $p_\theta(a,c|z)$, another MLP takes as input the latent sample $z$, and outputs the concatenation of reconstructed embeddings \smash{$\mathbf{e}_\theta^j(z)$} and click probabilities \smash{$p_\theta^{j,c}(c_j | z)$} for each slot $j$ in the slate. We then derive logits for the item probabilities \smash{$p_\theta^{j, a}(a_j | z)$} by taking the dot-product of the reconstructed embedding \smash{$\mathbf{e}_\theta^j(z)$} with the embeddings of all items in the collection. For collection items, we use the current version of embeddings learned within the encoder, but we prevent the gradient from back-propagating to them using the stop-gradient operator to avoid potential degenerate solutions.

In summary, the VAE is pre-trained by maximizing the ELBO on the task of reconstructing slates and corresponding clicks, i.e., by minimizing
$\mathcal{L}_{\theta, \phi}^{\mathrm{GeMS}} = \mathbb{E}_{a,c \sim \mathcal{D}} [ \mathcal{L}_{\theta, \phi}^{\mathrm{GeMS}}(a,c)]$ with:
\begin{equation}
    \begin{split}
        \mathcal{L}_{\theta, \phi}^{\mathrm{GeMS}}(a,c) 
        ={}&  \overset{\textup{slate reconstruction}}{\overbrace{\sum_{j = 1}^k \log p_\theta^{j,a}(a_j | z_\phi(a,c))}} + {}\\[-1.5mm]
        &\lambda \overset{\textup{click reconstruction}}{\overbrace{\sum_{j = 1}^k \log p_\theta^{j,c}(c_j | z_\phi(a,c))}} +{}  \\[-1.5mm] 
        &\beta \overset{\textup{KL-divergence}}{\overbrace{\sum_{i=1}^d \left(\sigma_{\phi, i}^2 + \mu_{\phi, i}^2 - \log \sigma_{\phi, i} - 1 \right)}}
    \end{split}
\label{eq:loss}
\end{equation}

\vspace*{-0.5mm}\noindent%
where \smash{$z_\phi(a,c) = \mu_\phi(a,c) + \text{diag}(\sigma^2_\phi(a,c)) \cdot \epsilon$}, for $\epsilon \sim \mathcal{N}(\mathbf{0},\mathbf{I})$. Here, $d$ is the dimension of the latent space, and $\beta$ and $\lambda$ are hyperparameters controlling the respective weight of the KL term and the click reconstruction term. Note that the KL term takes this simple form due to the Gaussian assumption on $q_\phi(z |a, c)$ and the $\mathcal{N}(\mathbf{0}, \mathbf{I})$ prior.


\subsection{RL agent and belief encoder}
\label{rl-framework}

After the pre-training step described in Section \ref{gems}, the parameters of GeMS are frozen and we use its decoder as the ranker in our RL framework. The RL agent can then be trained to maximize the discounted return by taking proto-actions within the VAE's latent space. To generate a slate \smash{$(i^1, \dots, i^k)$} from the agent's proto-action $z$, we take for each slot $j \in \{1,\ldots,k\}$ the most likely item according to the decoder: \smash{$ i^j = \mathrm{arg}\max_{i \in \mathcal{I}} p_\phi^{j,a}(i | z)$}.

Since our focus within the RL framework is on the choice of the ranker, we adopt a standard implementation of the belief encoder and the agent: the former is modeled by a gated recurrent unit (GRU) \citep{Cho2014} taking as input the concatenation of item embeddings and respective clicks from each slate, and the latter is a soft actor-critic (SAC) \citep{SAC} algorithm. We chose SAC because it is a well-established RL algorithm, known for its strong performance and data-efficiency in continuous control. Additionally, SAC adds an entropy term incentivizing exploration which we have noticed during our experiments to be important to attain high performance in highly stochastic recommendation environments. 

\section{Baselines and their assumptions}
\label{baselines}

We evaluate our proposed method against four main baselines derived from prior work. In this section, we describe these baselines as well the assumptions on user behavior that they formulate in order to make the combinatorial problem of slate recommendation tractable. By doing so, we are able to compare the assumptions made by these baselines and highlight the generality of our method in Table \ref{tab:assumptions}.
Note that we only report from previous studies the mechanism used for slate generation, which is the topic of this study, and ignore other design choices. 

\header{SoftMax} In \citep{youtube-topk,IRecGAN}, the authors reduce the combinatorial problem of slate optimization to the simpler problem of item optimization: the policy network output is a softmax layer over all items in the collection, and items are sampled with replacement to form slates.\\
Doing so requires the mild assumption that \textit{the Q-value of the slate can be linearly decomposed into item-specific Q-values} (\textbf{DQ}). But more importantly, it also requires two strong assumptions, namely
    \textit{users can click on at most one item per slate} (\textbf{1CL}) and 
    \textit{the returns of items in the same slate are mutually independent} (\textbf{MI}). 
Together, these assumptions are restrictive, because their conjunction means that the click probability of an item in the slate does not depend on the item itself. Indeed, having dependent click probabilities (to enforce the single click) and independent items in the slate is compatible only if click probabilities do not depend on items.

\header{SlateQ} \citet{slateQ} propose a model-based approach in which the click behavior of the user is given, and Q-learning \citep{q-learning} is used to plan 
and approximate users' dynamic preferences. On top of the earlier DQ and 1CL, it requires \textit{access to the true relevance and click model} (\textbf{CM}), which is an unfair advantage compared to other methods. For computational efficiency reasons, we adopt the faster variant referred to as QL-TT-TS in the original paper.

\begin{table}[t]
    \centering
    \caption{Comparison of assumptions made by prior work. Our method only requires access to logged interaction data.}
    \vspace{0.1cm}
    \begin{adjustbox}{center}
    \begin{tabular}{l cccccccc}
    \toprule
         & \textbf{1CL} & \textbf{DQ} & \textbf{MI} & \textbf{CM} & \textbf{SP} & \textbf{EIB} & \textbf{LD}  \\
         \midrule
        SoftMax \cite{youtube-topk, IRecGAN} & \cmark & \cmark & \cmark & \xmark & \xmark & \xmark & \xmark \\
        SlateQ \cite{slateQ} & \cmark & \cmark & \xmark & \cmark & \xmark  & \xmark & \xmark \\
        WkNN \cite{wolpertinger} & \xmark & \cmark & \xmark & \xmark & \cmark & \cmark & \cmark \\
        TopK & \xmark & \xmark & \xmark & \xmark & \cmark & \xmark & \cmark \\
        GeMS (Ours) & \xmark & \xmark & \xmark & \xmark & \xmark & \xmark & \cmark \\
        \bottomrule
    \end{tabular}
    \end{adjustbox}
    \label{tab:assumptions}
\end{table}

\header{TopK} Even though, to the best of our knowledge, no work has proposed this approach, we include it in our set of baselines as it is a natural way to deal with slate recommendation. The agent takes continuous actions in the space of item embeddings, and we generate slates by taking the $k$ items from the collection with the closest embeddings to the action, according to a similarity metric (the dot-product in practice). This method therefore assumes the \textit{availability of logged data of past interactions} (\textbf{LD}), in order to pre-train item embeddings. In our experiments, we evaluate two variants of this baseline: \textit{TopK (MF)}, where item embeddings are learned by matrix factorization \cite{Koren2009}, and \textit{TopK (ideal)}, which uses ideal item embeddings, i.e., the embeddings used internally by the simulator (see Section \ref{simulator}). The latter version clearly has an unfair advantage. 
Also, because ranking items this way assumes that the most rewarding items should appear on top, it makes the \textit{sequential presentation} (\textbf{SP}) assumption from \citep{wolpertinger} that \emph{the true click model is top-down and fading}, i.e., if $c(i)$ indicates that item $i$ has been clicked and $l \leqslant k$ is the position of $i$ in slate $a$, then $P(c(i)|s, a) = P(c(i)|s, a_{\leqslant l}) \leqslant P(c(i)|s, \tilde{a}_{\leqslant l-1})$, where $a_{\leqslant l} = (i^1, \dots, i^{l-1}, i)$ and $\tilde{a}_{\leqslant l-1} = (i^1, \dots, i^{l-2}, i)$.

\vspace{0.1cm}

\header{WkNN} In \citep{wolpertinger}, the authors propose a finer-grained and potentially more capable variant of TopK referred to as \textit{Wolpertinger} \citep{DulacArnold2015}: the agent takes actions in the product-space of item embeddings over slate slots, i.e., continuous actions of dimension $k \times d$, where $d$ is the dimension of item embeddings. Then, for each slot in the slate, $p$ candidate items are selected by Euclidean distance with embeddings of items from the collection, and every candidate item's contribution to the Q-value is evaluated in a greedy fashion. Besides LD and DQ, WkNN requires two strong assumptions to ensure submodularity of the Q-function: sequential presentation SP and \textit{execution is best} (\textbf{EIB}), i.e., \textit{recommendations that are risky on the short term are never worth it}. Formally, this translates as: $\mathbb{P}(R(s, \pi_1(s)) = 0) \geqslant \mathbb{P}(R(s, \pi_2(s)) = 0) \Rightarrow V^{\pi_1}(s) \leqslant V^{\pi_2}(s)$ for any policies $\pi_1$, $\pi_2$.
Note that it partly defeats the purpose of long-term optimization.

\vspace{0.2cm}\noindent%
In Table \ref{tab:assumptions}, we summarize the assumptions made by each baseline. In comparison to prior work, our proposed framework has a single assumption: the availability of logged data with slates and associated clicks (LD), as Table~\ref{tab:assumptions} indicates. This assumption is by no means restrictive as such logged data is readily available in common industrial recommendation settings.

On top of these baselines, we also include a \textbf{random} policy and a \textbf{short-term oracle} as reference points. The short-term oracle has access to the true user and item embeddings, enabling it to select the items with the highest relevance probability in each slate. Therefore, at each turn of interaction, it gives an upper bound on the immediate reward but it is unable to cope with boredom and influence phenomena.

\section{Experimental Setup}
\label{experimental-setup}

\begin{table*}[t]
    \centering
    \caption{Average cumulative number of clicks on the test set for our 6 simulated environments. Bold: best method; underlined: 2$^\text{nd}$-best method; $^\dagger$: statistically significantly better than all other methods. 95\% confidence intervals are given in parentheses. Methods grouped under ``Disclosed env.'' have access to privileged information about the environment and can therefore not be fairly compared with ``Undisclosed env.'' methods.}
    \vspace{0.2cm}
    \begin{adjustbox}{center}
    \scalebox{1.0}{
    \begin{tabular}{c r@{ \ }l r@{ \ }l r@{ \ }l r@{ \ }l r@{ \ }l r@{ \ }l}
    \toprule
        \multicolumn{1}{c}{ } & \multicolumn{6}{c}{\large{\textbf{Focused item embeddings}}} & \multicolumn{6}{c}{\large{\textbf{Diffuse item embeddings}}}\\
        \cmidrule(r){2-7}\cmidrule{8-13}
        \textbf{Method} 
        & \multicolumn{2}{c}{\textbf{TopDown}} 
        & \multicolumn{2}{c}{\textbf{Mixed}} 
        & \multicolumn{2}{c}{\textbf{DivPen}} 
        & \multicolumn{2}{c}{\textbf{TopDown}} 
        & \multicolumn{2}{c}{\textbf{Mixed}} 
        & \multicolumn{2}{c}{\textbf{DivPen}} \\
        \midrule
        \multirow{3}{*}{\rotatebox[origin=c]{90}{\begin{tabular}{@{}c@{}}\footnotesize\textbf{Disclosed}\\[-0.15cm]\footnotesize\textbf{env.}\end{tabular}} $\left\lbrace\begin{array}{l}
                \text{Short-term oracle}\\
                \text{SAC+TopK (ideal)\phantom{X}} \\
                \text{SlateQ}
                \end{array}\right.\hspace{0.3cm}$} & $107.7 $ & & $101.6 $ & & $85.4 $ & & $96.7$ & & $94.6$ & & $78.8$ & \\
        & $429.0$ & $(\pm 5.9)$ 
        & $384.1$ & $(\pm 13.5)$ 
        & $386.3$ & $(\pm 15.5)$ 
        & $373.9$ & $(\pm 25.0)$ 
        & $371.9$ & $(\pm 36.4)$ 
        & $341.3$ & $(\pm 55.3)$ \\
        & $206.5$ & $(\pm 4.1)$ 
        & $202.7$ & $(\pm 3.4)$ 
        & $119.0$ & $(\pm 3.9)$ 
        & $209.5$ & $(\pm 5.4)$ 
        & $192.7$ & $(\pm 5.1)$ 
        & $117.8$ & $(\pm 5.8)$ \\
        \midrule
        \multirow{5}{*}{\rotatebox[origin=c]{90}{\begin{tabular}{@{}c@{}}\footnotesize\textbf{Undisclosed}\\[-0.15cm]\footnotesize\textbf{env.}\end{tabular}}\hspace{0.1cm}$\left\lbrace\begin{array}{l}
                \text{Random}\\
                \text{REINFORCE+SoftMax}\\
                \text{SAC+WkNN}\\
                \text{SAC+TopK (MF)}\\
                \text{SAC+GeMS (Ours)}\\
                \end{array}\right.$}  
                & $33.8$ & $(\pm 0.2)$ 
                & $33.9$ & $(\pm 0.2)$ 
                & $33.6$ & $(\pm 0.2)$ 
                & $33.3$ & $(\pm 0.2)$ 
                & $33.2$ & $(\pm 0.2)$ 
                & $32.9$ & $(\pm 0.2)$ \\
        & $248.1$ & $(\pm 19.3)$ 
        & $\underline{233.5}$ & $(\pm 18.5)$ 
        & $\underline{249.1}$ & $(\pm 11.6)$ 
        & $249.5$ & $(\pm 15.3)$ 
        & $\underline{214.7}$ & $(\pm 25.0)$ 
        & $213.8$ & $(\pm 27.1)$ \\
        & $98.5$ & $(\pm 8.9)$ 
        & $97.7$ & $(\pm 10.8)$ 
        & $95.5$ & $(\pm 9.9)$ 
        & $107.2$ & $(\pm 8.9)$ 
        & $89.8$ & $(\pm 7.4)$ 
        & $92.5$ & $(\pm 5.0)$ \\
        & $\underline{254.4}$ & $(\pm 17.1)$ 
        & $232.7$ & $(\pm 19.4)$ 
        & $242.2$ & $(\pm 15.4)$ 
        & $\underline{249.7}$ & $(\pm 10.3)$ 
        & $184.1$ & $(\pm 1.3)$ 
        & $\underline{231.4}$ & $(\pm 13.3)$ \\
        & $ \mathbf{305.3}$\rlap{$^\dagger$} & $\mathbf{(\pm 21.9)}$ 
        & $\mathbf{242.6}$ & $\mathbf{(\pm 21.5)}$ 
        & $\mathbf{254.1}$ & $\mathbf{(\pm 27.7)}$ 
        & $\mathbf{300.0}$\rlap{$^\dagger$} & $\mathbf{(\pm 42.8})$ 
        & $\mathbf{260.6}$\rlap{$^\dagger$} & $\mathbf{(\pm 27.2)}$ 
        & $\mathbf{249.6}$ & $(\pm 37.6)$ \\
        \bottomrule
    \end{tabular}
    }
    \end{adjustbox}
    \label{tab:main_res}
\end{table*}

\subsection{Simulator}
\label{simulator}

We design a simulator that allows us to observe the effect of lifting the assumptions required by the baselines, and we experiment with several simulator variants to ensure generalizability. We summarize our main design choices below and refer the reader to our code available online\footnote{\href{https://naver/github/gems}{https://naver/github/gems}} for a more detailed description.


\header{Item and user embeddings}
Following scenario \textbf{(S)}, our simulator includes $1,000$ items. We consider a cold-start situation where users are generated on-the-fly for each new trajectory. Items and users are randomly assigned embeddings of size $20$, corresponding to ten $2$-dimensional topics: $\mathbf{e} = (\mathbf{e}^1, \dots, \mathbf{e}^{10})$. Each $2$-dimensional vector $\mathbf{e}^t$ is meant to capture the existence of subtopics within topic $t$.
The embedding of a user or item $x$ is generated using the following process: \begin{enumerate*}[label=(\roman*)]\item sample topic propensities $w_x^t \sim \mathcal{U}(0, 1)$ and normalize such that $\sum_t w_x^t = 1$; \item sample topic-specific components $\mathbf{\epsilon}_x^t \sim \mathcal{N}(\mathbf{0}, 0.4 \cdot \mathbf{I}_2)$ and rescale as $\mathbf{e}^t_x = w_x^t \cdot \min(|\mathbf{\epsilon}_x^t|,1))$; and \item normalize the embedding $\mathbf{e}_x = (\mathbf{e}_x^1, \ldots, \mathbf{e}_x^{10})$ such that $\|\mathbf{e}_x\| = 1$. \end{enumerate*} Each item is associated to a main topic, defined as $t(i) = \mathrm{arg}\max_{1 \leqslant t \leqslant 10} \| \mathbf{e}_i^t \| $. 

To accomodate different types of content and platforms, we derive two variants of item embeddings in the simulator: one with embeddings obtained as described above, and one with embeddings for which we square and re-normalize each component. In Section~ \ref{experiments}, we highlight this difference in peakedness by referring to the former as \textit{diffuse embeddings} and the latter as \textit{focused embeddings}.

\header{Relevance computation}
The relevance probability of item $i$ for user $u$ is a monotonically increasing function of the dot-product between their respective embeddings: \smash{$\mathrm{rel}(i, u) = \sigma ( {\mathbf{e}_i}^T \mathbf{e}_u )$},
where $\sigma$ is a sigmoid function.

\header{Boredom and influence effects}
User embeddings can be affected by two mechanisms: \textit{boredom} and \textit{influence}. Each item $i$ clicked by user $u$ influences the user embedding in the next interaction turn as: $ \mathbf{e}_u \leftarrow \omega \mathbf{e}_u + (1 - \omega) \mathbf{e}_i $, where we set $\omega = 0.9$ in practice. Additionally, if in the last $10$ items clicked by user $u$ five have the same main topic $t^b$, then $u$ gets bored with this topic, meaning we put \smash{$\mathbf{e}_u^{t^b} = \mathbf{0}$} for $5$ turns. These mechanisms have been defined to penalize myopic behavior and encourage long-term strategies.

\header{Click model}
Users click on recommended items according to a position-based model, i.e., the click probability is the product of item-specific attractiveness and rank-specific examination probabilities: $ \mathbb{P}(c |i, r) = A_i \times E_r $. Specifically, we define for an item located at rank $r$: \smash{$E_r = \nu \varepsilon^r + (1 - \nu) \varepsilon^{k+1-r}$} with $\varepsilon = 0.85$. It is a mixture of the terms \smash{$\varepsilon^r$} and \smash{$\varepsilon^{k+1-r}$}, which respectively capture the top-down and bottom-up browsing behaviors. We use two variants of this click model in our experiments: \textit{TopDown} with $\nu = 1.0$ and \textit{Mixed} with $\nu = 0.5$. The attractiveness of an item is set to its relevance in TopDown and Mixed. In addition, we consider a third variant \textit{DivPen} which also penalizes slates that lack diversity: $A_i$ is down-weighted by a factor of $3$ if more than $4$ items from the slate have the same main topic (as in Mixed, we also set $\nu = 0.5$ for DivPen). 

\vspace*{1mm}\noindent%
In summary, our experiments are performed on 6 simulator variants defined by the choice of  item embedding peakedness (\textit{diffuse item embeddings} or \textit{focused item embeddings}) and the choice of  click model (\textit{TopDown}, \textit{Mixed}, or \textit{DivPen}).

\subsection{Implementation and evaluation details}
\label{implementation}

Our implementation aims to be as standard as possible, considering the literature on RL, in order to ensure reproducibility. All baselines are paired with SAC \citep{SAC}, except SlateQ which is based on Q-Learning \citep{q-learning}, and SoftMax, which we pair with REINFORCE \citep{REINFORCE} because it requires a discrete action space and a discretized variant of SAC led to lower performance in our experiments. We implement all agents using two-layer neural networks as function approximators, and use target networks for Q-functions in Slate-Q and SAC. For hyperparameters common to baselines and our method, we first performed a grid search over likely regions of the space on baselines, and re-used the selected values for our method. For all methods we use the Adam optimizer with learning rates of $0.001$ for Q-networks and $0.003$ for policy networks when applicable, as well as a discount factor $\gamma = 0.8$ and a polyak averaging parameter $\tau = 0.002$. For the hyperparameters specific to our method ($d$, $\beta$ and $\lambda$), we perform a grid search on the TopDown environment with focused item embeddings and select the combination with the highest validation return. This combination is then re-used on all other environments. The searched ranges were defined as $d \in \{16, 32\}$, $\beta \in \{0.1, 0.2, 0.5, 1.0, 2.0\}$ and $\lambda \in \{0.0, 0.2, 0.5, 1.0\}$.

For methods making the (LD) assumption, we generated a dataset of $100$K user trajectories (with $100$ interactions turns each) from an $\epsilon$-greedy oracle policy with $\epsilon = 0.5$, i.e., each recommended item is selected either uniformly randomly or by an oracle, with equal probabilities. The VAE in GeMS is trained on this dataset for $10$ epochs with a batch size of $256$ and a learning rate of $0.001$. For approaches requiring pre-trained item embeddings (TopK and WkNN), we learn a simple matrix factorization model on the generated dataset by considering as positive samples the pairs composed of the user in the trajectory and each clicked item in their recommended slates.


In all of our experiments, we compare average cumulative rewards over 10 seeded runs, corresponding to ten initializations of the agent's parameters. In the case of GeMS, the seed also controls the initialization of the VAE model during pre-training. We train agents for $100$K steps. Each step corresponds to a user trajectory, composed of $100$ interaction turns (i.e., $100$ slates successively presented to the user) for a unique user. Every $1,000$ training steps, we also evaluate the agents on $200$ validation user trajectories. Finally, the agents are tested by selecting the checkpoint with the highest validation return and applying it on $500$ test user trajectories. Confidence intervals use Student's $t$-distribution, and statistical tests are Welch's $t$-test. Both are based on a $95\%$ confidence level.

\begin{figure*}[t]
\centering
    \begin{subfigure}{.32\textwidth}
    \centering
    \includegraphics[width = 1.15\textwidth]{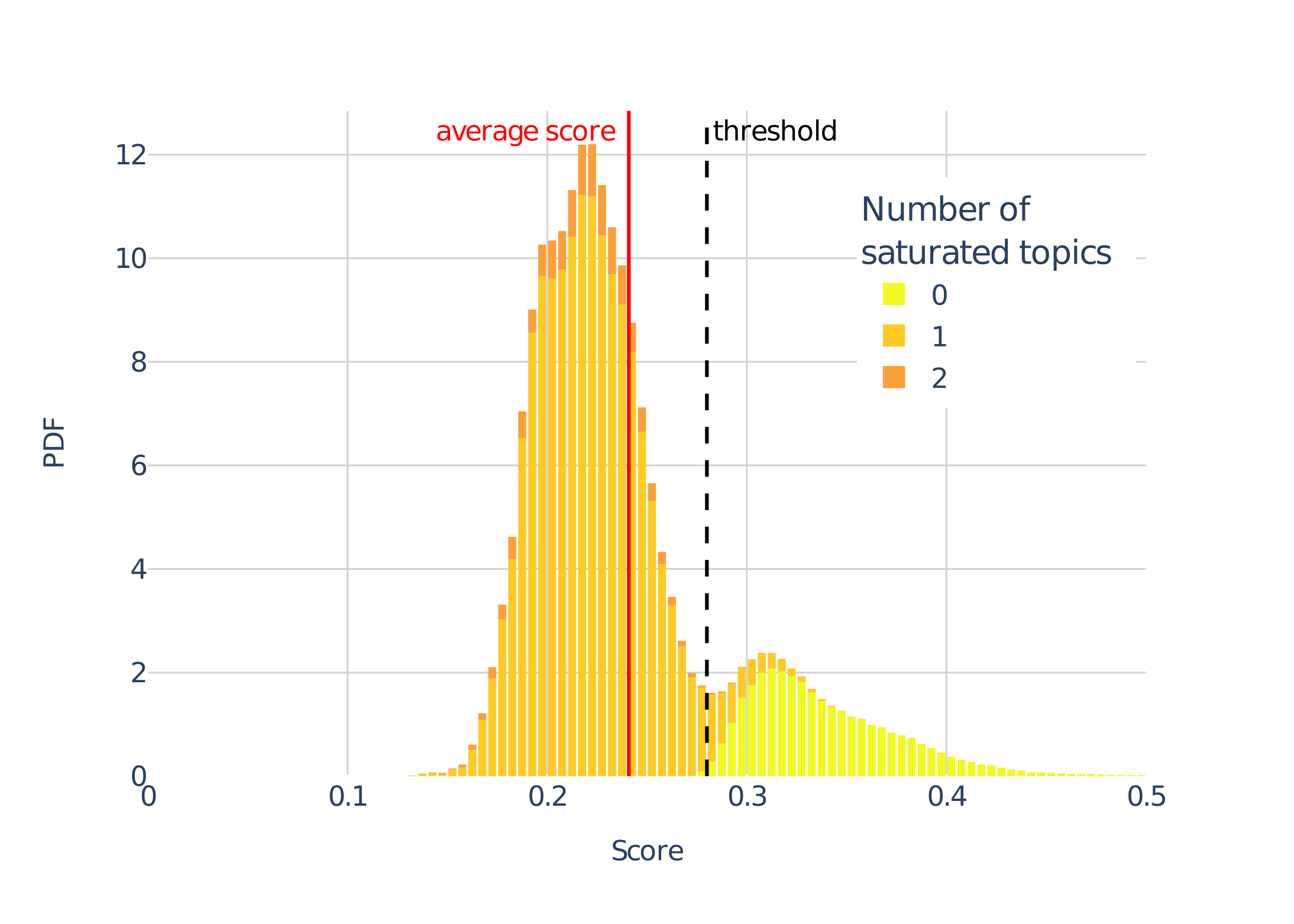} 
    \vspace{-0.8cm}
    \caption{Short-term oracle.}
    \label{fig:div_sto}
    \end{subfigure}
    \hfill
    \begin{subfigure}{.32\textwidth}
    \centering
    \includegraphics[width = 1.15\textwidth]{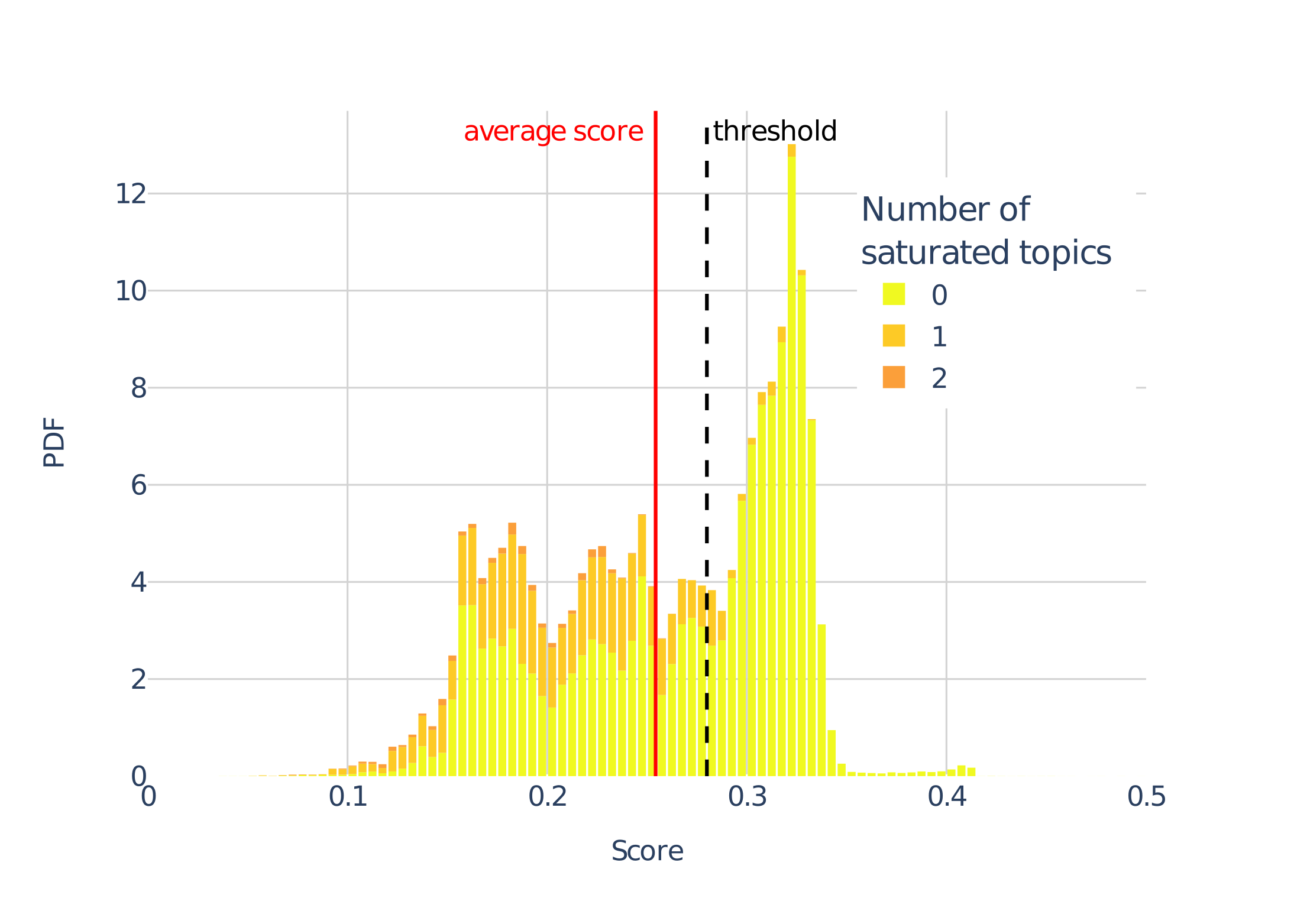} 
    \vspace{-0.8cm}
    \caption{SAC+GeMS with $\gamma = 0$.}
    \label{fig:div_gamma0}
    \end{subfigure}
    \hfill
    \begin{subfigure}{.35\textwidth}
    \centering
    \includegraphics[width = 1.05\textwidth]{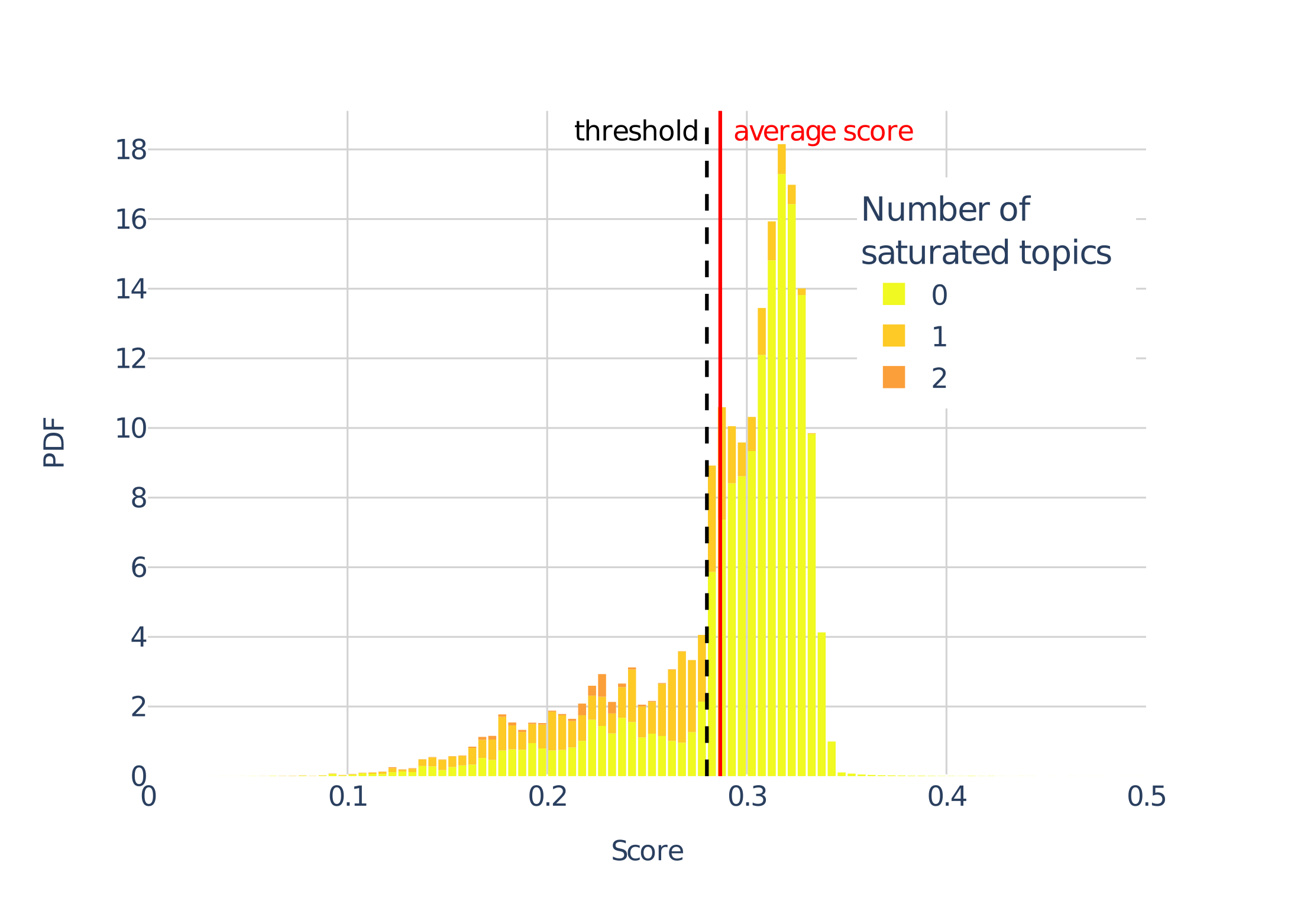} 
    \vspace{-0.8cm}
    \caption{SAC+GeMS with $\gamma = 0.8$.}
    \label{fig:div_gamma0.8}
    \end{subfigure}
\vspace{0.2cm}
\caption[Distribution of the relevance scores of items recommended by a short-term oracle, SAC+GeMS with $\gamma = 0$ and SAC+GeMS with $\gamma = 0.8$. The myopic approaches (left, center) lead to more boredom than the long-term approach (right).]{Distribution of the relevance scores of items recommended by \begin{enumerate*}[label=(\alph*)] \item a short-term oracle, \item SAC+GeMS with $\gamma = 0$ and \item SAC+GeMS with $\gamma = 0.8$ \end{enumerate*}. Boredom penalizes item scores and is visualized by orange areas. The myopic approaches (left, center) lead to more boredom than the long-term approach (right), and therefore to lower average item scores (solid red lines).}
\label{fig:short-vs-long-term}       
\end{figure*} 

\section{Results}
\label{experiments}

In our experiments, we investigate the following research questions:
\begin{enumerate*}[label=(\textbf{RQ\arabic*})]
    \item How does our slate recommendation framework based on GeMS compare to previous methods when the underlying assumptions of the latter are lifted?
    \item Does the proposed GeMS framework effectively balance immediate and future rewards to avoid boredom?
    \item How do the balancing hyperparameters $\beta$ and $\lambda$ in GeMS impact the downstream RL performance?
\end{enumerate*}

\subsection{Comparison of our method against baselines (RQ1)}
\label{exp:main-res}

In this section, we compare the performance of our method and baselines on a wide array of simulated environments, corresponding to the six environments described in Section \ref{simulator}.

\header{Overview of the results} Table \ref{tab:main_res} shows the average test return (i.e., cumulated reward or cumulated number of clicks) after training on $100$K user trajectories. We group methods into two categories: \textit{Disclosed env.}, i.e., methods leveraging hidden environment information, and \textit{Undisclosed env.}, i.e., methods that consider the environment as a black-box and are therefore practically applicable. A first observation we can draw, regardless of the specific environment used, is that the short-term oracle is easily beaten by most approaches. Indeed, the simulator penalizes short-sighted recommendations that lead to boredom: in these environments, \emph{diversity is required to reach higher returns.}
We can also observe the superiority of SAC+TopK (Ideal). This is not surprising, as this method benefits from an unfair advantage~-- access to true item embeddings~-- but it suggests that practically applicable methods could be augmented with domain knowledge to improve their performance. However, despite having access to privileged information, SlateQ's performance is subpar, especially in DivPen environments. Its lower performance might be explained by its approximate optimization strategy and restrictive single-click assumption.

\header{Overall comparison of methods} \emph{The proposed SAC+GeMS compares favorably to baselines across the range of environments we simulate.} Out of the 6 tested environments, SAC+GeMS obtained the best average results on all of them, among which 3 show a statistically significant improvement over all other methods. SAC+WkNN performs very poorly: we hypothesize that the approach suffers from the curse of dimensionality due to the larger action space (200 dimensions in our experiments) and the assumption made by the approach that candidate items need to be close to target item embeddings according to the Euclidean distance. SAC+TopK (MF) is more competitive, but the large difference with SAC+TopK (ideal) suggests that TopK is very sensitive to the quality of item embeddings. Despite its very restrictive assumptions and lack of theoretical guarantees in our setup, REINFORCE+SoftMax was a very competitive baseline overall. However, while its best checkpoint had high return, its training was unstable and failed to converge in our experiments, which suggests it may be unreliable. 

\header{Comparisons across environments} The TopDown environment is the easiest for most methods, regardless of the type of item embeddings. This is not surprising as all methods besides Random either assume a top-down click model, sample items in a top-down fashion or rely on data from a top-down logging policy. However, it is worth noting that other factors can dominate the performance, such as sub-optimality of item embeddings for SAC+TopK (MF). 
Conversely, DivPen was harder for most methods, because it requires a strong additional constraint to obtain high returns: intra-slate diversity must be high. SAC+GeMS was also affected by these dynamics, but remained able to beat other methods by generating diverse slates. 
Finally, the use of diffused item embeddings does not appear to cause lower returns for GeMS, compared with focused ones, but is associated with larger confidence intervals for SAC+GeMS: indeed, pivot items spanning multiple topics are more likely to be attractive, at the expense of more fine-grained strategies, making the training process uncertain.

\subsection{GeMS overcomes boredom to improve its return (RQ2)}
\label{exp:long-term}

In Section~\ref{intro} we highlighted that long-term optimization with RL can penalize myopic behavior such as recommending only highly relevant but similar items, which may lead to boredom. In this section, we verify that SAC+GeMS is able to adapt its slate selection to cope with boredom. We recall that in our simulated environments (detailed in Section~\ref{simulator}), users get bored of a particular topic whenever $5$ of their latest $10$ clicks were on items from that topic. When a topic is saturated, its corresponding dimensions in the user embedding are set to $\mathbf{0}$, which has the effect of diminishing the attractiveness of future items presented to the user. It is therefore necessary to avoid boredom in order to reach higher returns, even if it comes at the cost of lower immediate rewards.

\begin{figure*}[t]
\centering
    \begin{subfigure}{.47\textwidth}
    \includegraphics[width = \textwidth]{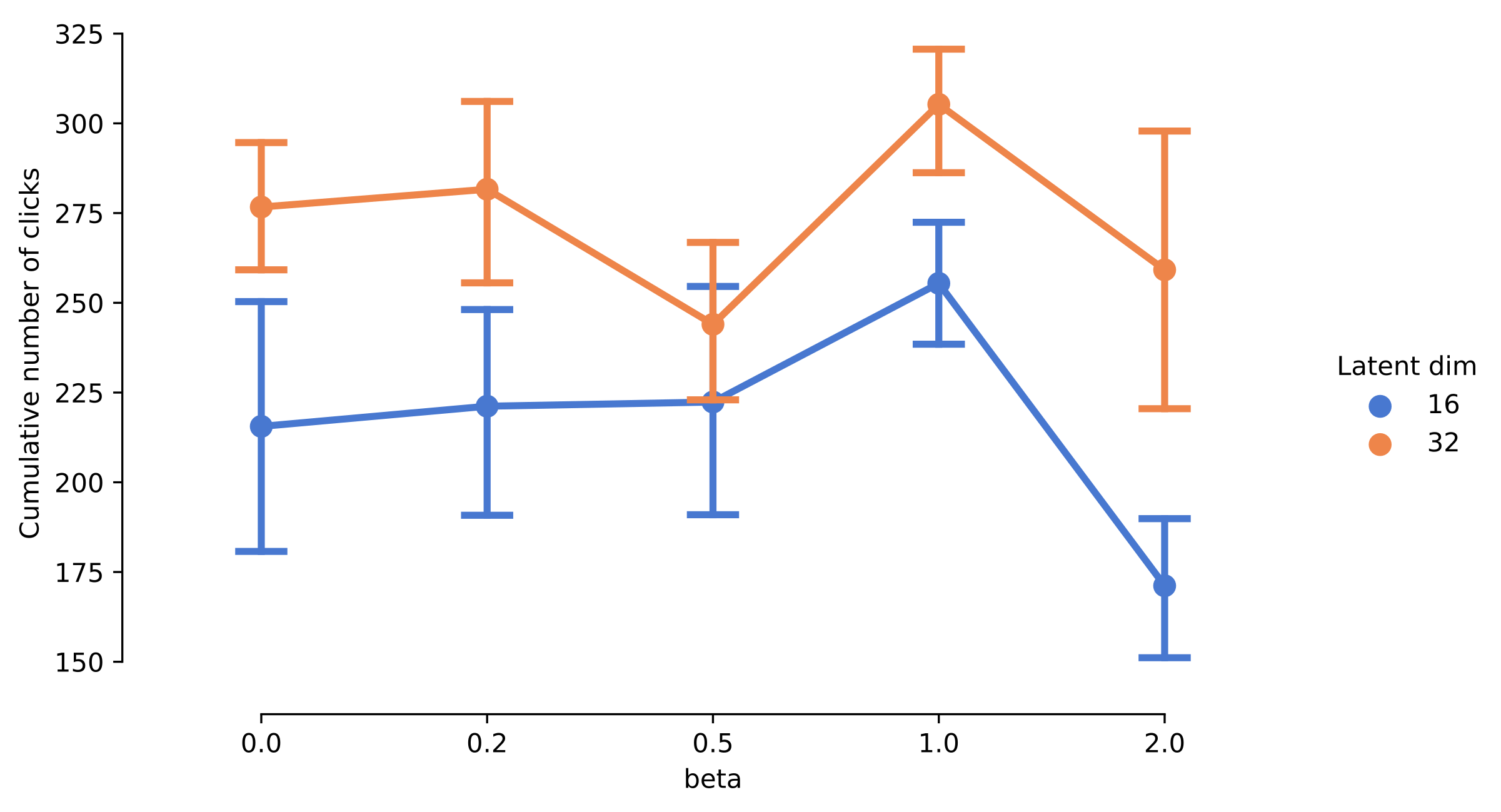} 
    \caption{Impact of $\beta$ for $\lambda = 0.5$.}
    \label{fig:beta-tradeoff}
    \end{subfigure}
    \hfill
    \begin{subfigure}{.47\textwidth}
    \includegraphics[width = \textwidth]{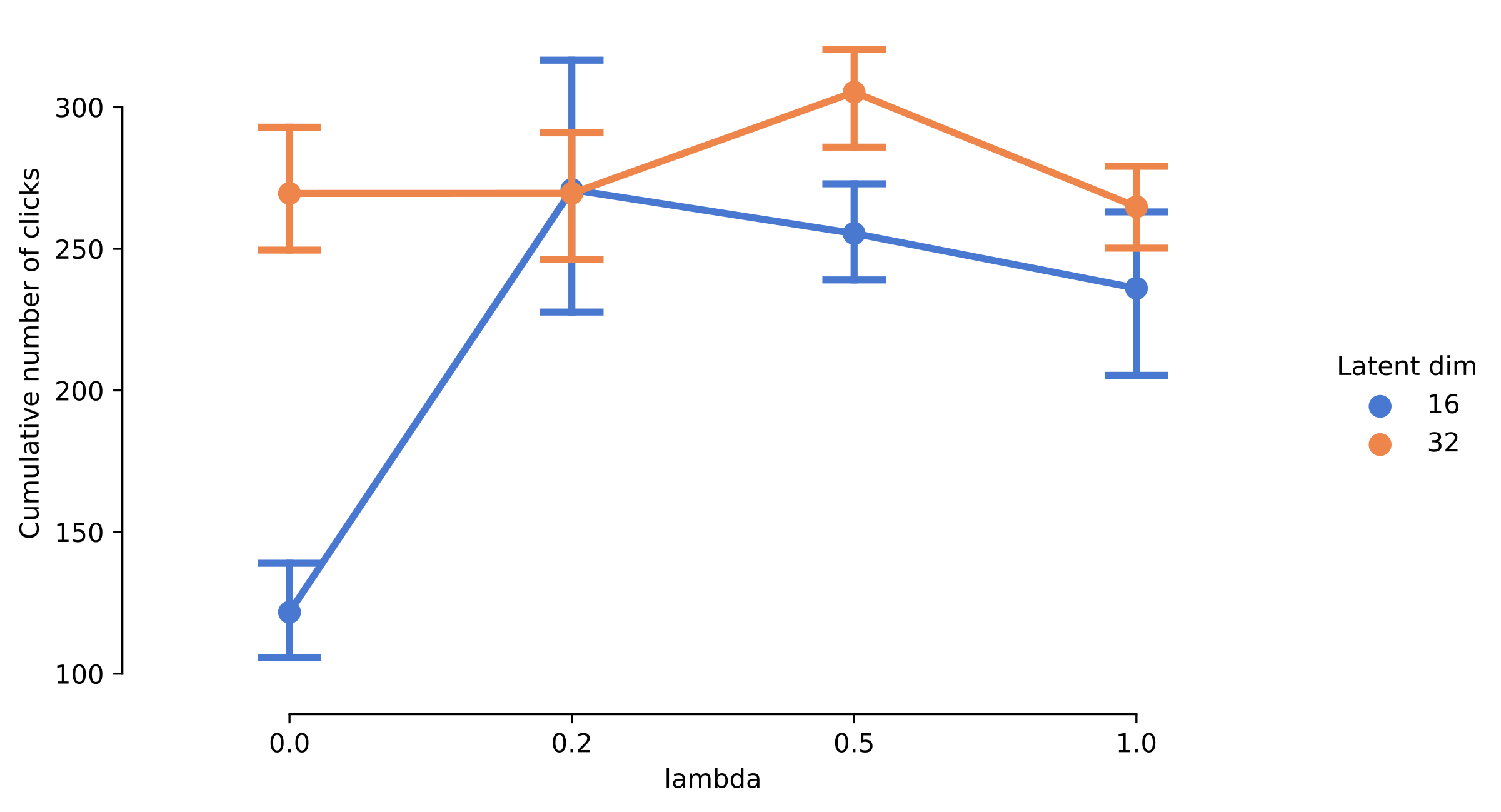} 
    \caption{Impact of $\lambda$ for $\beta = 1.0$.}
    \label{fig:lambda-tradeoff}
    \end{subfigure}
\vspace{0.2cm}
\caption{Average cumulative number of clicks on the validation set obtained by SAC+GeMS with its best validation checkpoint, for different values of $\beta$ and $\lambda$ (defined in Section \ref{gems}). We also display 95\% confidence intervals.}
\label{fig:tradeoffs}       
\end{figure*}

In this section, we compare three approaches on the TopDown environment with focused item embeddings: 
\begin{enumerate*}[label=(\roman*)]
\item the short-term oracle (STO) always maximizing the immediate reward, 
\item SAC+GeMS with $\gamma = 0.8$ (i.e., our proposed method) where $\gamma$ is the discount factor of the RL algorithm, and
\item SAC+GeMS with $\gamma = 0$ which does not explicitly include future rewards in its policy gradient. 
\end{enumerate*}
In this environment, SAC+GeMS$^{\gamma = 0.8}$ achieves an average test return of $305.3$, while SAC+GeMS$^{\gamma = 0}$ reaches $194.3$, and STO only obtains $107.7$. These results suggest that long-term optimization is indeed required to reach higher returns. It may seem surprising that SAC+GeMS$^{\gamma = 0}$ gets better returns than STO, but its training objective incentivizes \emph{average} immediate rewards, which implicitly encourages it to avoid low future rewards. However, adopting an explicit mechanism to account for its causal effect on the user (i.e., setting $\gamma = 0.8$) allows SAC+GeMS to improve its decision-making.

In Figure \ref{fig:short-vs-long-term}, we plot the distribution of item scores (i.e., the dot-product between internal user and item embeddings as defined in Section \ref{simulator}) for the items recommended in slates by each of the three methods, with the same seed for all three plots. The dashed vertical line shows the score threshold of $0.28$ needed to reach a relevance probability of $0.5$. Therefore, items on the left of this line have a lower click probability while items on the right have a higher click probability. The color indicates how many topics were saturated when the agent recommended that particular item whose score is plotted: one can see that when the user is bored of at least one topic, items become less attractive as scores are reduced. 

When no topic is saturated (i.e., yellow distribution), STO recommends items with excellent scores (above the threshold and up to $0.45$): as a consequence, STO gets high immediate rewards. However, by doing so it incurs a lot of boredom (large orange areas). Overall, it leads to lower expected scores (solid red line) and therefore fewer clicks.
Conversely, SAC+GeMS$^{\gamma = 0.8}$ sacrifices some immediate reward (yellow distribution shifted to the left) but causes very little boredom (small orange area). Overall, \emph{by trading off relevance and diversity, SAC+GeMS$^{\gamma = 0.8}$ yields good immediate rewards while limiting boredom.} It therefore gets higher average scores. SAC+GeMS$^{\gamma = 0}$ exhibits an intermediate behavior due to its limited capabilities: it recommends items of varying relevance, yet leads to substantial boredom (larger orange area than for $\gamma = 0.8$).



\subsection{Balancing hyperparameters $\beta$ and $\lambda$ (RQ3)}
\label{exp:tradeoffs}

In Section \ref{gems}, we suggested that the choice of $\beta$ and $\lambda$ leads to trade-offs that may impact the downstream performance of SAC+GeMS. As a reminder, $\beta$ adjusts the importance of accurate reconstruction versus smoothness and structure in the latent space (i.e., controllability), while $\lambda$ weights the click reconstruction with respect to the slate reconstruction. Next, we verify our intuition on the importance of these trade-offs by reporting (in Figure~\ref{fig:tradeoffs}) the best validation return obtained for different values of said hyperparameters, on the TopDown environment with focused item embeddings.

Figure \ref{fig:beta-tradeoff} suggests that, indeed, there exists a ``sweet spot'' in the selection of $\beta$. It confirms the intuition described in Section~\ref{gems} and the observation of \citet{Pivot-CVAE}: \emph{$\beta$ must be appropriately balanced in order to ensure high performance on the downstream RL task.} Specifically, we found that choosing $\beta = 1.0$ leads to the highest return overall, regardless of whether a latent dimension of $16$ or $32$ is used.

The impact on the downstream performance of the trade-off between slate and click reconstruction (Figure \ref{fig:lambda-tradeoff}) is less prominent but can still be observed.
It justifies our choice to add the click reconstruction term in the loss (Eq. \ref{eq:loss}), even though clicks output by GeMS' decoder are not used during RL training. This also confirms the importance of introducing and adjusting the hyperparameter $\lambda$: \emph{modeling clicks jointly with slates improves the final performance of SAC+GeMS, but properly weighting the click reconstruction objective with respect to the slate reconstruction objective is necessary.}


\section{Conclusion}
\label{conclusion}

We have presented GeMS, a slate representation learning method based on variational auto-encoders for slate recommendation with reinforcement learning. This method has the notable advantage of being flexible, allowing full-slate modeling and lightweight assumptions, in contrast with existing approaches. 

\header{Findings and broader impact}
Our experiments across a wide array of environments demonstrate that GeMS compares favorably against existing slate representation methods in practical settings. Moreover, our empirical analysis highlights that it effectively balances immediate and future rewards, and that the trade-offs imposed by $\beta$ and $\lambda$ significantly impact the RL downstream performance, indicating that properly balancing these hyperparameters is critical.
Our work suggests that generative models are a promising direction for representing rich actions such as slates.

\header{Limitations} Our simulated experiments demonstrate the effectiveness of GeMS for representing slates in an RL framework. However, it is well-known that online training of RL agents is too expensive and risky, and that in practice agents must be trained offline, i.e., directly from logged data \citep{youtube-topk}. We did not address here the specific challenges of offline RL, as we wished to isolate the contribution of the slate representation to downstream performance.

\header{Future work}  In future work, we will investigate how generative models can be leveraged in the offline setting, in different scenarios, or with even richer actions. We also plan to look into improvements of the architectures used for structured action representations, for example by using domain knowledge and user models.

\begin{acks}
This research was (partially) funded by the Hybrid Intelligence Center, a 10-year program funded by the Dutch Ministry of Education, Culture and Science through the Netherlands Organisation for Scientific Research, \url{https://hybrid-intelligence-centre.nl}.
All content represents the opinion of the authors, which is not necessarily shared or endorsed by their respective employers and/or sponsors.
\end{acks}

\bibliographystyle{ACM-Reference-Format}
\balance
\bibliography{main}


\begin{thebibliography}{36}


\ifx \showCODEN    \undefined \def \showCODEN     #1{\unskip}     \fi
\ifx \showDOI      \undefined \def \showDOI       #1{#1}\fi
\ifx \showISBNx    \undefined \def \showISBNx     #1{\unskip}     \fi
\ifx \showISBNxiii \undefined \def \showISBNxiii  #1{\unskip}     \fi
\ifx \showISSN     \undefined \def \showISSN      #1{\unskip}     \fi
\ifx \showLCCN     \undefined \def \showLCCN      #1{\unskip}     \fi
\ifx \shownote     \undefined \def \shownote      #1{#1}          \fi
\ifx \showarticletitle \undefined \def \showarticletitle #1{#1}   \fi
\ifx \showURL      \undefined \def \showURL       {\relax}        \fi
\providecommand\bibfield[2]{#2}
\providecommand\bibinfo[2]{#2}
\providecommand\natexlab[1]{#1}
\providecommand\showeprint[2][]{arXiv:#2}

\bibitem[\protect\citeauthoryear{Anderson, Maystre, Anderson, Mehrotra, and
  Lalmas}{Anderson et~al\mbox{.}}{2020}]%
        {spotify-div}
\bibfield{author}{\bibinfo{person}{Ashton Anderson}, \bibinfo{person}{Lucas
  Maystre}, \bibinfo{person}{Ian Anderson}, \bibinfo{person}{Rishabh Mehrotra},
  {and} \bibinfo{person}{Mounia Lalmas}.} \bibinfo{year}{2020}\natexlab{}.
\newblock \showarticletitle{Algorithmic Effects on the Diversity of Consumption
  on Spotify}. In \bibinfo{booktitle}{\emph{WWW '20}}.
  \bibinfo{pages}{2155–2165}.
\newblock


\bibitem[\protect\citeauthoryear{Athey, Chetty, Imbens, and Kang}{Athey
  et~al\mbox{.}}{2019}]%
        {surrogate-index}
\bibfield{author}{\bibinfo{person}{Susan Athey}, \bibinfo{person}{Raj Chetty},
  \bibinfo{person}{Guido~W Imbens}, {and} \bibinfo{person}{Hyunseung Kang}.}
  \bibinfo{year}{2019}\natexlab{}.
\newblock \bibinfo{booktitle}{\emph{The Surrogate Index: Combining Short-Term
  Proxies to Estimate Long-Term Treatment Effects More Rapidly and Precisely}}.
\newblock \bibinfo{type}{{T}echnical {R}eport}. \bibinfo{institution}{National
  Bureau of Economic Research}.
\newblock


\bibitem[\protect\citeauthoryear{Bai, Guan, and Wang}{Bai
  et~al\mbox{.}}{2019}]%
        {IRecGAN}
\bibfield{author}{\bibinfo{person}{Xueying Bai}, \bibinfo{person}{Jian Guan},
  {and} \bibinfo{person}{Hongning Wang}.} \bibinfo{year}{2019}\natexlab{}.
\newblock \showarticletitle{A Model-Based Reinforcement Learning with
  Adversarial Training for Online Recommendation}. In
  \bibinfo{booktitle}{\emph{NeurIPS~'19}}. \bibinfo{pages}{10734--10745}.
\newblock


\bibitem[\protect\citeauthoryear{Bakshy, Messing, and Adamic}{Bakshy
  et~al\mbox{.}}{2015}]%
        {Bakshy2015}
\bibfield{author}{\bibinfo{person}{Eytan Bakshy}, \bibinfo{person}{Solomon
  Messing}, {and} \bibinfo{person}{Lada Adamic}.}
  \bibinfo{year}{2015}\natexlab{}.
\newblock \showarticletitle{{Exposure to Ideologically Diverse News and Opinion
  on Facebook}}.
\newblock \bibinfo{journal}{\emph{Science}} \bibinfo{volume}{348},
  \bibinfo{number}{6239} (\bibinfo{year}{2015}), \bibinfo{pages}{1130--1132}.
\newblock


\bibitem[\protect\citeauthoryear{Botteghi, Poel, Sirma{\c{c}}ek, and
  Brune}{Botteghi et~al\mbox{.}}{2021}]%
        {SA-WM}
\bibfield{author}{\bibinfo{person}{Nicol{\`{o}} Botteghi},
  \bibinfo{person}{Mannes Poel}, \bibinfo{person}{Beril Sirma{\c{c}}ek}, {and}
  \bibinfo{person}{Christoph Brune}.} \bibinfo{year}{2021}\natexlab{}.
\newblock \showarticletitle{Low-Dimensional State and Action Representation
  Learning with {MDP} Homomorphism Metrics}.
\newblock \bibinfo{journal}{\emph{arXiv:2107.01677}} (\bibinfo{year}{2021}).
\newblock


\bibitem[\protect\citeauthoryear{Chandak, Theocharous, Kostas, Jordan, and
  Thomas}{Chandak et~al\mbox{.}}{2019}]%
        {AC-RA}
\bibfield{author}{\bibinfo{person}{Yash Chandak}, \bibinfo{person}{Georgios
  Theocharous}, \bibinfo{person}{James Kostas}, \bibinfo{person}{Scott Jordan},
  {and} \bibinfo{person}{Philip Thomas}.} \bibinfo{year}{2019}\natexlab{}.
\newblock \showarticletitle{Learning Action Representations for Reinforcement
  Learning}. In \bibinfo{booktitle}{\emph{ICML '19}}.
  \bibinfo{pages}{941--950}.
\newblock


\bibitem[\protect\citeauthoryear{Chandar, St.~Thomas, Maystre, Pappu,
  Sanchis-Ojeda, Wu, Carterette, Lalmas, and Jebara}{Chandar
  et~al\mbox{.}}{2022}]%
        {survival-models}
\bibfield{author}{\bibinfo{person}{Praveen Chandar}, \bibinfo{person}{Brian
  St.~Thomas}, \bibinfo{person}{Lucas Maystre}, \bibinfo{person}{Vijay Pappu},
  \bibinfo{person}{Roberto Sanchis-Ojeda}, \bibinfo{person}{Tiffany Wu},
  \bibinfo{person}{Ben Carterette}, \bibinfo{person}{Mounia Lalmas}, {and}
  \bibinfo{person}{Tony Jebara}.} \bibinfo{year}{2022}\natexlab{}.
\newblock \showarticletitle{Using Survival Models to Estimate User Engagement
  in Online Experiments}. In \bibinfo{booktitle}{\emph{WWW '22}}.
  \bibinfo{pages}{3186–3195}.
\newblock


\bibitem[\protect\citeauthoryear{Chen, Beutel, Covington, Jain, Belletti, and
  Chi}{Chen et~al\mbox{.}}{2019a}]%
        {youtube-topk}
\bibfield{author}{\bibinfo{person}{Minmin Chen}, \bibinfo{person}{Alex Beutel},
  \bibinfo{person}{Paul Covington}, \bibinfo{person}{Sagar Jain},
  \bibinfo{person}{Francois Belletti}, {and} \bibinfo{person}{Ed~H. Chi}.}
  \bibinfo{year}{2019}\natexlab{a}.
\newblock \showarticletitle{Top-K Off-Policy Correction for a REINFORCE
  Recommender System}. In \bibinfo{booktitle}{\emph{WSDM '19}}.
  \bibinfo{pages}{456–464}.
\newblock


\bibitem[\protect\citeauthoryear{Chen, Li, Li, Jiang, Qi, and Song}{Chen
  et~al\mbox{.}}{2019b}]%
        {cascading-Q}
\bibfield{author}{\bibinfo{person}{Xinshi Chen}, \bibinfo{person}{Shuang Li},
  \bibinfo{person}{Hui Li}, \bibinfo{person}{Shaohua Jiang},
  \bibinfo{person}{Yuan Qi}, {and} \bibinfo{person}{Le Song}.}
  \bibinfo{year}{2019}\natexlab{b}.
\newblock \showarticletitle{Generative Adversarial User Model for Reinforcement
  Learning Based Recommendation System}. In \bibinfo{booktitle}{\emph{ICML
  '19}}. \bibinfo{pages}{1052--1061}.
\newblock


\bibitem[\protect\citeauthoryear{Cho, van Merrienboer, G{\"{u}}l{\c{c}}ehre,
  Bahdanau, Bougares, Schwenk, and Bengio}{Cho et~al\mbox{.}}{2014}]%
        {Cho2014}
\bibfield{author}{\bibinfo{person}{Kyunghyun Cho}, \bibinfo{person}{Bart van
  Merrienboer}, \bibinfo{person}{{\c{C}}aglar G{\"{u}}l{\c{c}}ehre},
  \bibinfo{person}{Dzmitry Bahdanau}, \bibinfo{person}{Fethi Bougares},
  \bibinfo{person}{Holger Schwenk}, {and} \bibinfo{person}{Yoshua Bengio}.}
  \bibinfo{year}{2014}\natexlab{}.
\newblock \showarticletitle{Learning Phrase Representations using {RNN}
  Encoder-Decoder for Statistical Machine Translation}. In
  \bibinfo{booktitle}{\emph{EMNLP '14}}. \bibinfo{pages}{1724--1734}.
\newblock


\bibitem[\protect\citeauthoryear{Dang, Bendersky, and Croft}{Dang
  et~al\mbox{.}}{2013}]%
        {VanDang13}
\bibfield{author}{\bibinfo{person}{Van Dang}, \bibinfo{person}{Michael
  Bendersky}, {and} \bibinfo{person}{W.~Bruce Croft}.}
  \bibinfo{year}{2013}\natexlab{}.
\newblock \showarticletitle{Two-Stage Learning to Rank for Information
  Retrieval}. In \bibinfo{booktitle}{\emph{ECIR '13}}.
  \bibinfo{pages}{423--434}.
\newblock


\bibitem[\protect\citeauthoryear{Dulac-Arnold, Evans, van Hasselt, Sunehag,
  Lillicrap, Hunt, Mann, Weber, Degris, and Coppin}{Dulac-Arnold
  et~al\mbox{.}}{2015}]%
        {DulacArnold2015}
\bibfield{author}{\bibinfo{person}{Gabriel Dulac-Arnold},
  \bibinfo{person}{Richard Evans}, \bibinfo{person}{Hado van Hasselt},
  \bibinfo{person}{Peter Sunehag}, \bibinfo{person}{Timothy Lillicrap},
  \bibinfo{person}{Jonathan Hunt}, \bibinfo{person}{Timothy Mann},
  \bibinfo{person}{Theophane Weber}, \bibinfo{person}{Thomas Degris}, {and}
  \bibinfo{person}{Ben Coppin}.} \bibinfo{year}{2015}\natexlab{}.
\newblock \showarticletitle{Deep Reinforcement Learning in Large Discrete
  Action Spaces}.
\newblock \bibinfo{journal}{\emph{arXiv:1512.07679}} (\bibinfo{year}{2015}).
\newblock


\bibitem[\protect\citeauthoryear{Flaxman, Goel, and Rao}{Flaxman
  et~al\mbox{.}}{2016}]%
        {Flaxman2016}
\bibfield{author}{\bibinfo{person}{Seth~R. Flaxman}, \bibinfo{person}{Sharad
  Goel}, {and} \bibinfo{person}{Justin~M. Rao}.}
  \bibinfo{year}{2016}\natexlab{}.
\newblock \showarticletitle{{Filter Bubbles, Echo Chambers, and Online News
  Consumption}}.
\newblock \bibinfo{journal}{\emph{Public Opinion Quarterly}}
  \bibinfo{volume}{80}, \bibinfo{number}{S1} (\bibinfo{year}{2016}),
  \bibinfo{pages}{298--320}.
\newblock


\bibitem[\protect\citeauthoryear{Ha and Schmidhuber}{Ha and
  Schmidhuber}{2018}]%
        {world_models}
\bibfield{author}{\bibinfo{person}{David Ha} {and} \bibinfo{person}{J\"{u}rgen
  Schmidhuber}.} \bibinfo{year}{2018}\natexlab{}.
\newblock \showarticletitle{Recurrent World Models Facilitate Policy
  Evolution}. In \bibinfo{booktitle}{\emph{NeurIPS '18}}.
  \bibinfo{pages}{2455--2467}.
\newblock


\bibitem[\protect\citeauthoryear{Haarnoja, Zhou, Abbeel, and Levine}{Haarnoja
  et~al\mbox{.}}{2018}]%
        {SAC}
\bibfield{author}{\bibinfo{person}{Tuomas Haarnoja}, \bibinfo{person}{Aurick
  Zhou}, \bibinfo{person}{Pieter Abbeel}, {and} \bibinfo{person}{Sergey
  Levine}.} \bibinfo{year}{2018}\natexlab{}.
\newblock \showarticletitle{Soft Actor-Critic: Off-Policy Maximum Entropy Deep
  Reinforcement Learning with a Stochastic Actor}. In
  \bibinfo{booktitle}{\emph{ICML '18}}. \bibinfo{pages}{1856--1865}.
\newblock


\bibitem[\protect\citeauthoryear{Hansen, Mehrotra, Hansen, Brost, Maystre, and
  Lalmas}{Hansen et~al\mbox{.}}{2021}]%
        {RL-diversity}
\bibfield{author}{\bibinfo{person}{Christian Hansen}, \bibinfo{person}{Rishabh
  Mehrotra}, \bibinfo{person}{Casper Hansen}, \bibinfo{person}{Brian Brost},
  \bibinfo{person}{Lucas Maystre}, {and} \bibinfo{person}{Mounia Lalmas}.}
  \bibinfo{year}{2021}\natexlab{}.
\newblock \showarticletitle{Shifting Consumption towards Diverse Content on
  Music Streaming Platforms}. In \bibinfo{booktitle}{\emph{WSDM '21}}.
  \bibinfo{pages}{238–246}.
\newblock


\bibitem[\protect\citeauthoryear{Hohnhold, O'Brien, and Tang}{Hohnhold
  et~al\mbox{.}}{2015}]%
        {long-term-better}
\bibfield{author}{\bibinfo{person}{Henning Hohnhold}, \bibinfo{person}{Deirdre
  O'Brien}, {and} \bibinfo{person}{Diane Tang}.}
  \bibinfo{year}{2015}\natexlab{}.
\newblock \showarticletitle{Focusing on the Long-Term: It's Good for Users and
  Business}. In \bibinfo{booktitle}{\emph{KDD '15}}.
  \bibinfo{pages}{1849–1858}.
\newblock


\bibitem[\protect\citeauthoryear{Ie, Jain, Wang, Narvekar, Agarwal, Wu, Cheng,
  Chandra, and Boutilier}{Ie et~al\mbox{.}}{2019}]%
        {slateQ}
\bibfield{author}{\bibinfo{person}{Eugene Ie}, \bibinfo{person}{Vihan Jain},
  \bibinfo{person}{Jing Wang}, \bibinfo{person}{Sanmit Narvekar},
  \bibinfo{person}{Ritesh Agarwal}, \bibinfo{person}{Rui Wu},
  \bibinfo{person}{Heng-Tze Cheng}, \bibinfo{person}{Tushar Chandra}, {and}
  \bibinfo{person}{Craig Boutilier}.} \bibinfo{year}{2019}\natexlab{}.
\newblock \showarticletitle{SlateQ: A Tractable Decomposition for Reinforcement
  Learning with Recommendation Sets}. In \bibinfo{booktitle}{\emph{IJCAI~'19}}.
  \bibinfo{pages}{2592--2599}.
\newblock


\bibitem[\protect\citeauthoryear{Jannach, Pu, Ricci, and Zanker}{Jannach
  et~al\mbox{.}}{2021}]%
        {Jannach2021}
\bibfield{author}{\bibinfo{person}{Dietmar Jannach}, \bibinfo{person}{Pearl
  Pu}, \bibinfo{person}{Francesco Ricci}, {and} \bibinfo{person}{Markus
  Zanker}.} \bibinfo{year}{2021}\natexlab{}.
\newblock \showarticletitle{Recommender Systems: Past, Present, Future}.
\newblock \bibinfo{journal}{\emph{{AI} Mag.}} \bibinfo{volume}{42},
  \bibinfo{number}{3} (\bibinfo{year}{2021}), \bibinfo{pages}{3--6}.
\newblock


\bibitem[\protect\citeauthoryear{Jiang, Gowal, Qian, Mann, and Rezende}{Jiang
  et~al\mbox{.}}{2019}]%
        {List-CVAE}
\bibfield{author}{\bibinfo{person}{Ray Jiang}, \bibinfo{person}{Sven Gowal},
  \bibinfo{person}{Yuqiu Qian}, \bibinfo{person}{Timothy~A. Mann}, {and}
  \bibinfo{person}{Danilo~J. Rezende}.} \bibinfo{year}{2019}\natexlab{}.
\newblock \showarticletitle{Beyond Greedy Ranking: Slate Optimization via
  List-CVAE}. In \bibinfo{booktitle}{\emph{ICLR '19}}.
\newblock


\bibitem[\protect\citeauthoryear{Kaelbling, Littman, and Cassandra}{Kaelbling
  et~al\mbox{.}}{1998}]%
        {belief}
\bibfield{author}{\bibinfo{person}{Leslie~Pack Kaelbling},
  \bibinfo{person}{Michael~L. Littman}, {and} \bibinfo{person}{Anthony~R.
  Cassandra}.} \bibinfo{year}{1998}\natexlab{}.
\newblock \showarticletitle{Planning and Acting in Partially Observable
  Stochastic Domains}.
\newblock \bibinfo{journal}{\emph{Artificial Intelligence}}
  \bibinfo{volume}{101}, \bibinfo{number}{1} (\bibinfo{year}{1998}),
  \bibinfo{pages}{99--134}.
\newblock


\bibitem[\protect\citeauthoryear{Kingma and Welling}{Kingma and
  Welling}{2014}]%
        {VAE}
\bibfield{author}{\bibinfo{person}{Diederik Kingma} {and} \bibinfo{person}{Max
  Welling}.} \bibinfo{year}{2014}\natexlab{}.
\newblock \showarticletitle{Auto-Encoding Variational Bayes}. In
  \bibinfo{booktitle}{\emph{ICLR '14}}.
\newblock


\bibitem[\protect\citeauthoryear{Koren, Bell, and Volinsky}{Koren
  et~al\mbox{.}}{2009}]%
        {Koren2009}
\bibfield{author}{\bibinfo{person}{Yehuda Koren}, \bibinfo{person}{Robert~M.
  Bell}, {and} \bibinfo{person}{Chris Volinsky}.}
  \bibinfo{year}{2009}\natexlab{}.
\newblock \showarticletitle{{Matrix Factorization Techniques for Recommender
  Systems}}.
\newblock \bibinfo{journal}{\emph{Computer}} \bibinfo{volume}{42},
  \bibinfo{number}{8} (\bibinfo{year}{2009}), \bibinfo{pages}{30--37}.
\newblock


\bibitem[\protect\citeauthoryear{Kullback and Leibler}{Kullback and
  Leibler}{1951}]%
        {KL}
\bibfield{author}{\bibinfo{person}{Solomon Kullback} {and}
  \bibinfo{person}{Richard~A. Leibler}.} \bibinfo{year}{1951}\natexlab{}.
\newblock \showarticletitle{{On Information and Sufficiency}}.
\newblock \bibinfo{journal}{\emph{The Annals of Mathematical Statistics}}
  \bibinfo{volume}{22}, \bibinfo{number}{1} (\bibinfo{year}{1951}),
  \bibinfo{pages}{79--86}.
\newblock


\bibitem[\protect\citeauthoryear{Liu, Sun, Ge, Pei, and Zhang}{Liu
  et~al\mbox{.}}{2021}]%
        {Pivot-CVAE}
\bibfield{author}{\bibinfo{person}{Shuchang Liu}, \bibinfo{person}{Fei Sun},
  \bibinfo{person}{Yingqiang Ge}, \bibinfo{person}{Changhua Pei}, {and}
  \bibinfo{person}{Yongfeng Zhang}.} \bibinfo{year}{2021}\natexlab{}.
\newblock \showarticletitle{Variation Control and Evaluation for Generative
  Slate Recommendations}. In \bibinfo{booktitle}{\emph{WWW '21}}.
  \bibinfo{pages}{436–448}.
\newblock


\bibitem[\protect\citeauthoryear{Masrour, Wilson, Yan, Tan, and
  Esfahanian}{Masrour et~al\mbox{.}}{2020}]%
        {Masrour2020}
\bibfield{author}{\bibinfo{person}{Farzan Masrour}, \bibinfo{person}{Tyler
  Wilson}, \bibinfo{person}{Heng Yan}, \bibinfo{person}{Pang{-}Ning Tan}, {and}
  \bibinfo{person}{Abdol{-}Hossein Esfahanian}.}
  \bibinfo{year}{2020}\natexlab{}.
\newblock \showarticletitle{Bursting the Filter Bubble: Fairness-Aware Network
  Link Prediction}. In \bibinfo{booktitle}{\emph{AAAI '20}}.
  \bibinfo{pages}{841--848}.
\newblock


\bibitem[\protect\citeauthoryear{McInerney, Brost, Chandar, Mehrotra, and
  Carterette}{McInerney et~al\mbox{.}}{2020}]%
        {RIPS}
\bibfield{author}{\bibinfo{person}{James McInerney}, \bibinfo{person}{Brian
  Brost}, \bibinfo{person}{Praveen Chandar}, \bibinfo{person}{Rishabh
  Mehrotra}, {and} \bibinfo{person}{Benjamin Carterette}.}
  \bibinfo{year}{2020}\natexlab{}.
\newblock \showarticletitle{Counterfactual Evaluation of Slate Recommendations
  with Sequential Reward Interactions}. In \bibinfo{booktitle}{\emph{KDD '20}}.
  \bibinfo{pages}{1779--1788}.
\newblock


\bibitem[\protect\citeauthoryear{Pariser}{Pariser}{2011}]%
        {Pariser2011}
\bibfield{author}{\bibinfo{person}{Eli Pariser}.}
  \bibinfo{year}{2011}\natexlab{}.
\newblock \bibinfo{booktitle}{\emph{{The Filter Bubble: What the Internet Is
  Hiding from You}}}.
\newblock \bibinfo{publisher}{The Penguin Press}.
\newblock


\bibitem[\protect\citeauthoryear{Rossi, Polderman, and Frasca}{Rossi
  et~al\mbox{.}}{2021}]%
        {frasca-closedloop}
\bibfield{author}{\bibinfo{person}{Wilbert~Samuel Rossi},
  \bibinfo{person}{Jan~Willem Polderman}, {and} \bibinfo{person}{Paolo
  Frasca}.} \bibinfo{year}{2021}\natexlab{}.
\newblock \showarticletitle{The Closed Loop between Opinion Formation and
  Personalised Recommendations}.
\newblock \bibinfo{journal}{\emph{IEEE Transactions on Control of Network
  Systems}} (\bibinfo{year}{2021}).
\newblock


\bibitem[\protect\citeauthoryear{Stooke, Lee, Abbeel, and Laskin}{Stooke
  et~al\mbox{.}}{2021}]%
        {representation-RL}
\bibfield{author}{\bibinfo{person}{Adam Stooke}, \bibinfo{person}{Kimin Lee},
  \bibinfo{person}{Pieter Abbeel}, {and} \bibinfo{person}{Michael Laskin}.}
  \bibinfo{year}{2021}\natexlab{}.
\newblock \showarticletitle{Decoupling Representation Learning from
  Reinforcement Learning}. In \bibinfo{booktitle}{\emph{ICML '21}}.
  \bibinfo{pages}{9870--9879}.
\newblock


\bibitem[\protect\citeauthoryear{Sunehag, Evans, Dulac{-}Arnold, Zwols,
  Visentin, and Coppin}{Sunehag et~al\mbox{.}}{2015}]%
        {wolpertinger}
\bibfield{author}{\bibinfo{person}{Peter Sunehag}, \bibinfo{person}{Richard
  Evans}, \bibinfo{person}{Gabriel Dulac{-}Arnold}, \bibinfo{person}{Yori
  Zwols}, \bibinfo{person}{Daniel Visentin}, {and} \bibinfo{person}{Ben
  Coppin}.} \bibinfo{year}{2015}\natexlab{}.
\newblock \showarticletitle{Deep Reinforcement Learning with Attention for
  Slate Markov Decision Processes with High-Dimensional States and Actions}.
\newblock \bibinfo{journal}{\emph{arXiv:1512.01124}} (\bibinfo{year}{2015}).
\newblock


\bibitem[\protect\citeauthoryear{Sutton and Barto}{Sutton and Barto}{2018}]%
        {REINFORCE}
\bibfield{author}{\bibinfo{person}{Richard Sutton} {and}
  \bibinfo{person}{Andrew Barto}.} \bibinfo{year}{2018}\natexlab{}.
\newblock \bibinfo{booktitle}{\emph{Reinforcement Learning: An Introduction}}.
\newblock \bibinfo{publisher}{MIT Press}, \bibinfo{pages}{326--329}.
\newblock


\bibitem[\protect\citeauthoryear{Waller and Anderson}{Waller and
  Anderson}{2019}]%
        {spotify-GS}
\bibfield{author}{\bibinfo{person}{Isaac Waller} {and} \bibinfo{person}{Ashton
  Anderson}.} \bibinfo{year}{2019}\natexlab{}.
\newblock \showarticletitle{Generalists and Specialists: Using Community
  Embeddings to Quantify Activity Diversity in Online Platforms}. In
  \bibinfo{booktitle}{\emph{WWW '19}}. \bibinfo{pages}{1954–1964}.
\newblock


\bibitem[\protect\citeauthoryear{Watkins and Dayan}{Watkins and Dayan}{1992}]%
        {q-learning}
\bibfield{author}{\bibinfo{person}{Christopher Watkins} {and}
  \bibinfo{person}{Peter Dayan}.} \bibinfo{year}{1992}\natexlab{}.
\newblock \showarticletitle{Q-learning}.
\newblock \bibinfo{journal}{\emph{Machine Learning}} \bibinfo{number}{8}
  (\bibinfo{year}{1992}), \bibinfo{pages}{279--292}.
\newblock


\bibitem[\protect\citeauthoryear{Zhou, Bajracharya, and Held}{Zhou
  et~al\mbox{.}}{2020}]%
        {PLAS}
\bibfield{author}{\bibinfo{person}{Wenxuan Zhou}, \bibinfo{person}{Sujay
  Bajracharya}, {and} \bibinfo{person}{David Held}.}
  \bibinfo{year}{2020}\natexlab{}.
\newblock \showarticletitle{{PLAS:} Latent Action Space for Offline
  Reinforcement Learning}. In \bibinfo{booktitle}{\emph{CoRL '20}}.
  \bibinfo{pages}{1719--1735}.
\newblock


\bibitem[\protect\citeauthoryear{Zou, Xia, Ding, Song, Liu, and Yin}{Zou
  et~al\mbox{.}}{2019}]%
        {rl-longterm}
\bibfield{author}{\bibinfo{person}{Lixin Zou}, \bibinfo{person}{Long Xia},
  \bibinfo{person}{Zhuoye Ding}, \bibinfo{person}{Jiaxing Song},
  \bibinfo{person}{Weidong Liu}, {and} \bibinfo{person}{Dawei Yin}.}
  \bibinfo{year}{2019}\natexlab{}.
\newblock \showarticletitle{Reinforcement Learning to Optimize Long-Term User
  Engagement in Recommender Systems}. In \bibinfo{booktitle}{\emph{KDD '19}}.
  \bibinfo{pages}{2810--2818}.
\newblock


\end{thebibliography}


\end{document}